\documentclass[twoside]{article}
\usepackage{fleqn,espcrc2}

\usepackage{graphicx}

\title{New Solutions of the T-Matrix Theory of the Attractive Hubbard Model}

\author{
K.S.D. Beach and R.J. Gooding\address{Dept. of Physics, Queen's University,
        Kingston, Ontario Canada K7L 3N6}%
        ,~~~and~
        F. Marsiglio\address{Dept. of Physics, University of Alberta 
        Edmonton, Alberta Canada T6G 2J1}}
       
\begin{document}

\begin{abstract}
We present a novel solution method with which we obtain the spectral
functions of the attractive Hubbard model in two dimensions in
a non self-consistent formulation of the T-matrix approximation. We use 
a partial fraction decomposition of the self energy from which we obtain 
the single-particle spectral functions to a remarkably high accuracy. Here we 
find the poles and residues of the self energy to a relative accuracy of 
$10^{-80}$ for an $8\times 8$ lattice, and plot the resulting density of states.
Our results show pseudogap physics as $T_c$ is approached from above; these analytical 
results are in agreement with recent Quantum Monte Carlo data.  
\vspace{1pc}
\end{abstract}

\maketitle

One can motivate a study of the attractive Hubbard model (AHM) by suggesting 
that it might provide a rudimentary description of the correlated electrons
in the high-$T_c$ materials.  One feature that theorists are keen to see 
reproduced in any model is the appearance of a normal-state pseudogap \cite{timusk} 
above $T_c$ that evolves smoothly into the energy gap of the superconducting
state. For example, QMC studies~\cite{Tremblay-QMC} have produced 
evidence of pseudogap physics in the two-dimensional AHM.

If the T-matrix approximation applied to the AHM can be shown to suppress the 
density of
states (DOS) at the Fermi level for temperatures above $T_c$, this would lend credence 
to the notion that the pseudogap is the result of strong pairing fluctuations.
In this brief report, we demonstrate that one can calculate the DOS
to an arbitrarily high accuracy for the non self-consistent (NSC) version of 
the T-matrix approximation, and that one in fact obtains results very similar
to the QMC data. We stress that this calculational method should be applicable
to any non self-consistent microscopic theory involving meromorphic functions.

To begin, we must calculate the fully interacting DOS within the
NSC T-matrix approximation.  Such a calculation is made possible by
the fact that we have access to the exact solution of
the NSC pair propagator, viz.\hfill\break 
${G}^{\rm pp}_{+-}(\vec{Q},i\nu_n)
= \chi(\vec{Q},i\nu_n)/(1-|U|\chi(\vec{Q},i\nu_n))$ where
the analytic continuation of $\chi(\vec{Q},i\nu_n)$ to the complex plane
is given by
\begin{equation} \label{EQ:continued-pair-suscept}
\bar{\chi}(\vec{Q},z) = \frac{1}{N} \sum_{\vec{k}} \frac{f[\xi_{\vec{k}}]
+ f[\xi_{\vec{Q}-\vec{k}}] - 1}{z - \xi_{\vec{k}} - \xi_{\vec{Q}-\vec{k}}}\:.
\end{equation}
The latter function is meromorphic with a finite number of simple poles on the
real axis (as usual, $f$ is the Fermi-Dirac distribution and $\xi$ is
the non-interacting single-particle energy relative to the chemical potential).
Likewise, the analytic continuation of the pair propagator,
\begin{equation} \label{EQ:continued-NSC-pair-prop}
\bar{G}^{\rm pp}_{+-}(\vec{Q},z) = 
  \frac{\frac{1}{N} \sum_{\vec{k}}
     \frac{f[\xi_{\vec{k}}]+f[\xi_{\vec{Q}-\vec{k}}]-1}
     {z - \xi_{\vec{k}} - \xi_{\vec{Q}-\vec{k}}}}
     {1-\frac{|U|}{N} \sum_{\vec{k}}
     \frac{f[\xi_{\vec{k}}]+f[\xi_{\vec{Q}-\vec{k}}]-1}
     {z - \xi_{\vec{k}} - \xi_{\vec{Q}-\vec{k}}}}\:,
\end{equation}
can be shown to be a meromorphic function with a finite number of simple poles.
This demonstrates that the pair propagator can be written as a rational polynomial 
with real coefficients, a consequence of which is that there are only two possibilities
for the placement of its singularities: they lie either on the
real axis (corresponding to the normal state) or in conjugate pairs
equally spaced above and below the real axis (corresponding to
the unstable state below $T_c$).
Moreover, since the degree of the polynomial in the denominator
is larger by one than that of the polynomial in the numerator,
the high-frequency asymptotic behaviour of the
pair propagator is governed by $\bar{G}^{\rm pp}_{+-}(\vec{Q},z)
\sim 1/z$ as $|z| \rightarrow \infty$.

Thus, in the normal state, the pair propagator admits a partial
fraction decomposition into a series of simple poles.  We write this as
\begin{eqnarray} \label{EQ:pair-prop-partial-frac}
\bar{G}^{\rm pp}_{+-}(\vec{Q},z) 
= -\frac{{\rm sgn}(E^{(1)}_{\vec{Q}})R^{(1)}_{\vec{Q}}}{z-E^{(1)}_{\vec{Q}}}&&\nonumber \\
-\frac{{\rm sgn}(E^{(2)}_{\vec{Q}})R^{(2)}_{\vec{Q}}}{z-E^{(2)}_{\vec{Q}}}-
\frac{{\rm sgn}(E^{(3)}_{\vec{Q}})R^{(3)}_{\vec{Q}}}{z-E^{(3)}_{\vec{Q}}}
- \dots&&
\end{eqnarray}
where the energies $E^{(l)}_{\vec{Q}}$ are real and
the residues $R^{(l)}_{\vec{Q}}>0$ are strictly positive.
[Below we denote the number of such poles for each $\vec{Q}$ component 
by $s_{\vec{Q}}$.] Quite simply, such a decomposition can be executed numerically 
in MapleV to arbitrary accuracy (without an associated increase
in computing time), and in what follows we used
{\bf Digits:=80} to obtain a relative accuracy of $10^{-80}$.

We now show how this result can be exploited to give us a practical,
numerical method for calculating the self-energy.  The first step
is to put the partial fraction form of the pair propagator,
Eq.~(\ref{EQ:pair-prop-partial-frac}), into the self energy
using the identity
$G^{\rm pp}_{+-}(\vec{Q},i\nu_n) = \bar{G}^{\rm pp}_{+-}(\vec{Q},z=i\nu_n)$
to recover the Matsubara frequency components.  Then, we complete
various Matsubara frequency sums, and the final result is
\begin{eqnarray}
\bar{\Sigma}(\vec{k},z) =~~~~~~~~~~~~~~~~~~~~~~~~~&&\\
\frac{U^2}{N}\sum_{\vec{Q}}\sum_{l=1}^{s_{\vec{Q}}}
\frac{{\rm sgn}(E^{(l)}_{\vec{Q}})R^{(l)}_{\vec{Q}}\bigl(
f[\xi_{\vec{Q}-\vec{k}}]+b[E^{(l)}_{\vec{Q}}]\bigr)}
{z+\xi_{\vec{Q}-\vec{k}}-E^{(l)}_{\vec{Q}}}\:&&\nonumber
\end{eqnarray}
where $b$ is the Bose distribution function.
Once the self-energy is calculated numerically by this procedure,
the Green's function follows from Dyson's equation, and the
single-particle spectral function follows in the usual way.
The DOS can then be constructed according to
\begin{equation}
{\cal N}(\omega) = -\frac{1}{\pi}\frac{2}{N}\sum_{\vec{k}}
\rm{Im} \bar{G}(\vec{k},\omega+i\eta)\:.
\end{equation}
This yields a DOS that is a series of $\delta$-function peaks;
when plotting our results we incorporate a small
artificial broadening to smooth the curves.

Above we show two figures for the resulting DOS for an $8\times 8$
two-dimensional square lattice for $|U|$ being half the band width. 
The chemical potential is varied to fix the density to be that
of a 1/4-filled system as the temperature is lowered. The pseudogap is
clearly visible, whereas at $T/t=1$ (not shown) the spectrum 
shows no anomaly at the Fermi level. 
Note that in this system (fixed density $\not=$ one)
$T_c=0$ \cite{prl}.

This work was supported by the NSERC of Canada.

\begin{figure}
\includegraphics{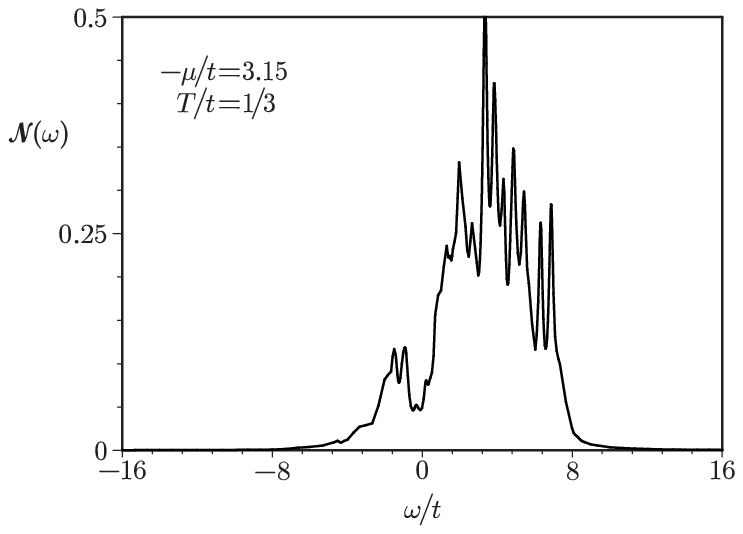}
\includegraphics{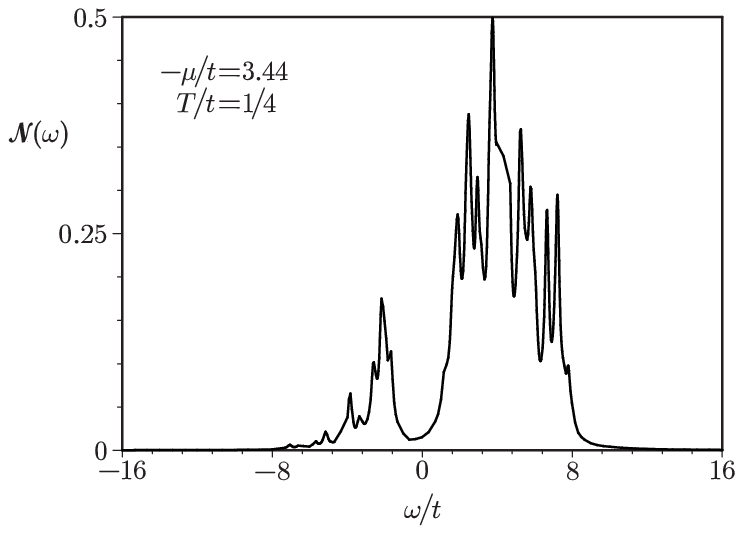}
\end{figure}

\end{document}